\documentclass[aps,twocolumn,pra,showpacs,floatfix]{revtex4-1}

\usepackage{amssymb}
\usepackage{amsmath}
\usepackage{graphicx}

\newcommand{\tm}{\tablenotemark}
\begin{document}

\title{\bf Atomic many-body effects and Lamb shifts in alkali metals}
\author{J. S. M. Ginges}
\affiliation{School of Physics, University of New South Wales,
Sydney NSW 2052, Australia}
\author{J. C. Berengut}
\affiliation{School of Physics, University of New South Wales,
Sydney NSW 2052, Australia}

\date{\today}

\begin{abstract}

We present a detailed study of the Flambaum-Ginges 
radiative potential method which enables the accurate inclusion of  
quantum electrodynamics (QED) radiative corrections in a simple manner in  
atoms, ions, and molecules over the range $10\le Z \le 120$, 
where $Z$ is the nuclear charge. 
Calculations are performed for binding energy shifts to the lowest
valence $s$, $p$, and $d$ waves over the series of alkali atoms Na to E119.
The high accuracy of the radiative potential method is demonstrated by
comparison with rigorous QED calculations in 
frozen atomic potentials, with deviations on the level of 1\%.
The many-body effects of core relaxation and second- and higher-order 
perturbation theory on the interaction of the valence electron with
the core are calculated. 
The inclusion of many-body effects tends to increase the size of the
shifts, with the enhancement particularly significant for $d$ waves; 
for K to E119, the self-energy shifts for $d$ waves are only an
order of magnitude smaller than the $s$-wave shifts.   
It is shown that account of many-body
effects is essential for an accurate description of the Lamb shift. 

\end{abstract}


\maketitle

\section{Introduction}

The increasingly accurate account of electron-electron correlations in
calculations of the properties of many-electron atoms, ions, and molecules has seen the need for 
quantum electrodynamics (QED) radiative corrections to be included in the formalism. 
The account of combined many-body effects and radiative corrections was
crucial in the atomic theory interpretation of the cesium parity violation
measurement \cite{wieman}, restoring an apparent deviation from the 
standard model of particle physics
\cite{Sush_ueh,JBS_ueh,KF2002,MST2002,SPVC2003,SPTY2005,rad_pot,review}. 
Tests of quantum electrodynamics in the measurements of transition 
frequencies in highly-charged many-electron ions also rely on an
accurate description of combined many-body and QED effects (see, e.g.,
\cite{blundell,CCJS}). The
increasing size of the radiative corrections with nuclear charge
$Z$ makes the account of such effects necessary in the accurate prediction
of the physical and chemical properties of the superheavy elements \cite{Schwerdt_review}.

The methods of rigorous (``exact'') QED that have had such great success in 
applications for single- or few-electron atoms and ions are not tractable for the
many-electron system \cite{mohr_review}. While it is possible to determine QED
corrections to atomic properties in the exact formalism in {\it frozen} atomic potentials (see,
e.g., Refs. \cite{labzowsky,sapcheng} for Lamb shifts to binding energies in
alkali atoms), in this approach important many-body effects such
as core relaxation and electron-electron correlations may be
prohibitively difficult to calculate. 

The Lamb shift is the physical radiative shift which is comprised, in the one-loop approximation, 
of the non-local self-energy and the local vacuum polarization shifts, the 
former giving the larger effect.
A number of approaches for estimating the Lamb shifts in
many-electron atoms in a simple manner have been put
forward, and we refer the reader to Ref. \cite{pyykko} for one 
such method and a description of earlier methods. 
Some approaches rely 
on rescaling the shifts, e.g., from the vacuum polarization.
Others involve the introduction of an approximate potential, which 
we term a ``radiative potential'', that mimics the self-energy effects. 
These potentials are found by fitting to the self-energy shifts for 
hydrogen-like ions.
A local radiative potential is appealing due to the ease in which it
may be included into many-body atomic or molecular computer codes with
the resulting full account of many-body effects. 

We introduced such a local radiative potential a decade ago 
in our work Ref. \cite{rad_pot}.  
This potential has been implemented in a number of calculations, including
in the calculation of the parity violating amplitude in Cs \cite{rad_pot} and other
atoms and ions \cite{roberts}, in the
spectra of heavy and superheavy atoms \cite{DDFG,radium,nature}, and in
highly-charged ions \cite{nandysahoo}. It has also been applied in Ref. \cite{thierfelder} with slightly
different fitting factors for the high-frequency part of the electric
potential with fitting to hydrogen-like $s$ waves for principal
quantum number $n=1$ to $n=5$.

Recent applications of other approaches include implementation \cite{chantler}
of a method based on Welton's idea \cite{IGD,welton}. 
See also Ref. \cite{tupbers,shabaev,dyall} for the development 
and application of non-local radiative potentials.

Shabaev, Tupitsyn, and Yerokhin have demonstrated the very high accuracy
of their non-local QED potential  \cite{shabaev,QEDMOD}, termed the ``model
operator'', by
comparing their results for self-energy shifts with those of exact  
QED \cite{sapcheng} performed in the same frozen atomic potentials. 
In their work, they hinted that the property of the non-locality
of the self-energy should be preserved for obtaining high accuracy.
This was based on the isolation of 
a term of the form $A_{\kappa}\exp{(-r/\alpha)}$, 
fitted to reproduce self-energy shifts for hydrogen-like ions for the lowest level in each wave $\kappa$,
where $\alpha$ is the fine-structure constant. (Throughout the paper we use atomic units,
$\hbar = e = m = 1$, $c=1/\alpha$.)
It was shown that this local potential 
yields $s$-wave self-energy shifts for neutral alkali atoms with increasingly large deviations
from the results of exact QED with increasing nuclear charge $Z$, the error for Fr
amounting to about $30\%$.

In the current work we determine the $s$-wave self-energy shifts to
binding energies in alkali atoms using the Flambaum-Ginges local
radiative potential \cite{rad_pot} in frozen atomic potentials. We demonstrate
that the accuracy of the radiative potential method is high, the
deviations from the results of exact QED 
on the level of $1\%$, the error roughly double that of
the model operator approach \cite{shabaev}.  
The simplicity of this potential (not much more complicated than the
Uehling potential) makes its inclusion into many-body methods and
codes straightforward. 

The combined self-energy and many-body effects on the binding energies in
neutral atoms have not been studied in detail before. In this work we
consider the many-body mechanisms and effects
of core relaxation and core-valence correlations on the self-energy
shifts of the neutral alkali atoms. Consideration of such many-body 
effects is crucial for obtaining the correct magnitude and sign of the 
shift for waves with orbital angular momentum $l>0$. 
We have introduced an $l$-dependence into the electric part of the
radiative potential which 
enables the $d$-level shifts to be controlled and the overall
accuracy of the potential improved. 
The many-body enhancement mechanisms that we have observed in this 
work for the self-energy are the same that we saw in our recent work on the vacuum 
polarization (Uehling) shifts \cite{GBUehling}.

\section{The radiative potential}

The Flambaum-Ginges radiative potential (FGRP) 
is a local potential that approximates the 
one-loop self-energy and vacuum polarization effects on 
electron energies and orbitals and may be readily included 
in many-body atomic structure 
calculations.
The derivation of this potential 
may be found in Ref. \cite{rad_pot}.
The self-energy part of the potential contains factors
that are found by fitting to self-energy shifts
for states of high principal quantum number for hydrogen-like ions.
In the current paper, we focus on the self-energy aspect of the
problem. We addressed
in detail many-body effects on the dominant  
contribution to the vacuum polarization (the Uehling potential) in our 
recent paper \cite{GBUehling}.

The following arguments justify the use of local radiative potentials
in neutral atoms: (i) the radiative QED interactions act at small
distances, on the order of the Compton wavelength $r\sim \alpha$,
where the electrons are unscreened by other electrons; (ii) the
binding energies of valence electrons in neutral
atoms are much smaller than the rest-mass energy, $\epsilon \sim 10^{-5}mc^2$. 
Therefore, in this unscreened region, the valence electrons in a neutral atom behave in
the same way as a weakly bound electron in a Coulomb potential. That
is, in this region, the wave functions of electrons in a neutral atom
$\varphi$ are proportional
to the electron wave functions $\varphi _{\rm H}$ of hydrogen-like ions with high
principal quantum number. Therefore, to good accuracy,
\begin{equation}
\label{eq:ratio}
\langle \varphi |V_{\rm SE}({\bf r},{\bf r}',\epsilon)|\varphi\rangle
= \langle \varphi_{\rm H} |V_{\rm SE}({\bf r},{\bf r}',\epsilon)|\varphi_{\rm H}\rangle \frac{\rho (r_n)}{\rho_{\rm H}(r_n)} \ ,
\end{equation}
where $\rho(r_n)=\varphi^{\dagger}(r_n)\varphi(r_n)$ is the density of
the electron wave function at the nucleus and the subscript ``${\rm H}$''
refers to the case for hydrogen-like ions.
Since the Uehling potential is localized in the nuclear vicinity, the
expression above may also be written as
\begin{equation}
\label{eq:ratiomethod}
\delta \epsilon_{\rm SE}
= \delta \epsilon_{\rm SE,H} \frac{\delta \epsilon_{\rm Ueh}}{\delta
  \epsilon _{\rm Ueh,H}} \ , 
\end{equation}
where $\delta \epsilon _{\rm SE}=\langle \varphi|V_{\rm
  SE}({\bf r},{\bf r}',\epsilon)|\varphi\rangle$ 
and $\delta \epsilon _{\rm Ueh}=\langle \varphi|V_{\rm
  Ueh}|\varphi\rangle$ are the self-energy and Uehling corrections to the
  binding energy.
This relation was used in Ref. \cite{pyykkoRatio} to estimate 
self-energy valence $s$-wave shifts to binding energies 
and later confirmed by rigorous self-energy
calculations in Ref. \cite{labzowsky}.
In a similar manner, based around Welton's idea of the fluctuating 
position of the electron,
where the dominant part of the self-energy shift for the $s$-waves  
is proportional to $\nabla ^2 V({\bf r})$ and $V({\bf r})$ is the
potential seen by the electron \cite{welton}, 
the self-energy shift may be approximated by the ratio
$\delta \epsilon_{\rm SE}
= \delta \epsilon_{\rm SE,H} \langle \varphi|\nabla ^2 V({\bf
    r})|\varphi\rangle/\langle \varphi_{\rm H}|\nabla ^2 V ({\bf
    r})|\varphi_{\rm H}\rangle$, Ref. \cite{IGD}. See also
Ref. \cite{chantler} for a recent implementation of this approach
and for other references. 
For states $l>0$, the ratio
$\delta \epsilon_{\rm SE}=\delta \epsilon_{\rm SE, H}\langle
\varphi|\beta \boldsymbol{\alpha} \cdot \nabla V({\bf
    r})|\varphi\rangle/\langle \varphi_{\rm H}|\beta
  \boldsymbol{\alpha}\cdot \nabla V ({\bf
    r})|\varphi_{\rm H}\rangle$
has been proposed \cite{ID1990}, where $\beta$ and $\boldsymbol{\alpha}$ are Dirac matrices.

Using the ratio methods above, one may yield reasonable estimates for the
self-energy corrections for valence $s$ orbitals of atoms and ions. 
However, for estimating shifts for orbitals with $l>0$, this procedure
may prove to be cumbersome or inadequate, since core relaxation corrections determine the size
and the sign of the effect \cite{derev_relax,rad_pot,GBUehling}.

The goal, then, is to extract from Eq.~(\ref{eq:ratio}) a local potential
that, when averaged over an orbital's wave function, gives the 
one-loop self-energy correction to the energy of the orbital.  
This potential may then be added to many-body atomic
structure codes in a simple manner.   
One may expect, from examination of Eq.~(\ref{eq:ratio}),
that as long as self-energy shifts for hydrogen-like ions are
reproduced with high accuracy by such a potential,
the accuracy for self-energy shifts for neutral atoms should also be
high. 

It is worth noting that the local potential that was isolated from the
non-local model operator considered in the 
work of Shabaev {\it et al.} \cite{shabaev} was fitted to the
(tightly-bound) 1$s$ state of hydrogen-like ions and then 
applied to the (loosely-bound) valence $s$ levels in neutral 
atoms. 
This is likely the reason for the deviations on the order of 10\% from
the results of exact QED for calculations 
in frozen atomic potentials.
Indeed, from Table VI in their work Ref. \cite{shabaev}, it is seen
that their local potential produces shifts for the $5s$ level in
hydrogen-like ions that deviate from the exact self-energy shifts by 
about 30\% for the heavier ions considered in that table, $40\le Z\le 92$.

\begin{table*}[bth]
\caption{Self-energy corrections to the binding
  energies for $s$, $p_{1/2}$, $p_{3/2}$, $d_{3/2}$, and $d_{5/2}$ states for hydrogen-like
  ions. Point-nucleus results of the radiative potential (FGRP) are compared
  with exact self-energy point-nucleus calculations \cite{mohrkim,highlse,shabaev}. 
The shifts are expressed as values of the function $F(Z\alpha)$, Eq. (\ref{eq:fzalpha}).}
\label{tab:hlike}
\begin{ruledtabular}
\begin{tabular}{llccccccccccc}
Ion & Z & $n$ & \multicolumn{9}{c}{$F(Z\alpha)$} \\
& & & \multicolumn{2}{c}{$ns$} &  \multicolumn{2}{c}{$np_{1/2}$} &  \multicolumn{2}{c}{$np_{3/2}$} & \multicolumn{2}{c}{$nd_{3/2}$} 
&  \multicolumn{2}{c}{$nd_{5/2}$}\\
&&&FGRP&Exact\tablenotemark[1]& FGRP&Exact\tablenotemark[1]&
FGRP&Exact\tablenotemark[1]
& FGRP&Exact\tablenotemark[1]& FGRP&Exact\tablenotemark[1]\\
\hline
Na &11&2 & 4.7878 & 4.6951&-0.0796&-0.1129&0.1656&0.1316&&&&\\
&&3 &4.7832 &4.7530&-0.0876&-0.0998& 0.1568&0.1434&-0.0416&-0.0426&0.0419&0.0409\\
&&4 &4.7808&4.7753&-0.0892&-0.0940&0.1548&0.1487&-0.0401&-0.0406&0.0434&0.0429\\
&&5 &4.7792&4.7860&-0.0898&-0.0908&0.1540&0.1516 & -0.0395&-0.0394&0.0440&0.0441\\
K&19 & 3 & 3.6825&3.6550 &-0.0628&-0.0790&0.1740&0.1555&-0.0409 &-0.0421 &0.0427&0.0416\\
&& 4 & 3.6784 &3.6754&-0.0643&-0.0721& 0.1718&0.1612&-0.0393&-0.0400&0.0444&0.0438\\
&& 5 & 3.6753 &3.6845&-0.0648&-0.0685&0.1709&0.1644& -0.0386&-0.0388 & 0.0451& 0.0450\\
Rb &37& 3 & 2.6315& 2.6041 &\ 0.0033& -0.0165 &0.2149  &0.1910 &-0.0389&-0.0401 &0.0455&0.0445\\
&&4 & 2.6220 &2.6186&\ 0.0024&-0.0066&0.2122&0.1982& -0.0367&-0.0376&0.0478&0.0472\\
&& 5 & 2.6144& 2.6227&\ 0.0022&-0.0015& 0.2110&0.2018&-0.0359&-0.0362&0.0487&0.0486\\
Cs &55&4&2.2172&2.2045&\ 0.0894&0.0805&0.2556&0.2431& -0.0325&-0.0334&0.0530&0.0523\\
&&5& 2.2027&2.2012 &\ 0.0891&\ 0.0867&0.2540&0.2475 & -0.0313&-0.0316&0.0542&0.0542\\
 &&6& 2.1915 && \ 0.0889&& 0.2531&&-0.0307&&0.0549&\\
Fr &87& 5& 2.0939& 2.0965&\ 0.3429&\ 0.3465&0.3438&0.3469&-0.0144&-0.0135&0.0700&0.0687\\
 & & 6& 2.0637 && \ 0.3388&&0.3419&&-0.0131&&0.0712&\\
 & & 7& 2.0404 && \ 0.3354&& 0.3404&&-0.0124&&0.0719&\\
E119 & 119 & 5 & 3.0499& 3.0642 &\ 1.1687&\ 1.257&0.4750&0.4686 & 0.0283 & 0.0311 & 0.0970 & 0.0880\\ 
& & 6 & 2.9407  & &\ 1.1299&  &0.4714&&0.0309&&0.0994&\\
& & 7 & 2.8596 &  &\ 1.1006& &0.4682& &0.0323&&0.1006&\\
& & 8 & 2.7975& &\ 1.0778 &  & 0.4654&&0.0331&&0.1012&\\  
\end{tabular}
\end{ruledtabular}
\tablenotetext[1]{Values found by interpolation of the exact calculations of Shabaev {\it et al.},
consistent with those of Mohr and Kim \cite{mohrkim} and Le Bigot {\it et al.} \cite{highlse}.}
\end{table*}

In the FGRP approach, the self-energy part of the radiative potential 
contains a magnetic formfactor term and an electric
formfactor term, divided into high- and low-frequency components,
\begin{equation}
V_{{\rm SE}}({\bf r})=V_{\rm mag}({\bf r}) +V_{\rm high}(r)+V_{\rm low}(r) \ .
\end{equation} 
For the point-nucleus case $V_{\rm nuc}^{\rm point}=Z/r$, the potentials
have the form \cite{rad_pot}
\begin{widetext}
\begin{eqnarray}
\label{eq:mag_point}
&&V_{\rm mag}^{\rm point} ({\bf r}) = \frac{i\alpha^2}{4\pi}\boldsymbol{\gamma}\cdot \boldsymbol{\nabla} \Big[
\Big(\frac{Z}{r}\Big)\Big( \int_1^{\infty}dt
\frac{1}{t^2\sqrt{t^2-1}}e^{-2tr/\alpha}-1\Big) \Big] \ , \\
\label{eq:high_point}
&&V_{\rm high}^{\rm point}(r)=-A_{l}(Z,r)\Big(\frac{\alpha}{\pi}\Big)\Big(\frac{Z}{r}\Big)
\int_{1}^{\infty}dt\frac{1}{\sqrt{t^2-1}}\Big[ \Big(
1-\frac{1}{2t^2}\Big)[\ln(t^2-1)+4\ln(1/Z\alpha +1/2)] 
-\frac{3}{2}+\frac{1}{t^2}\Big] e^{-2tr/\alpha} \ , \quad \\
\label{eq:low_point}
&&V_{\rm low}^{\rm point}(r)=-B_{l}(Z)Z^4\alpha^3e^{-Zr}  \ , 
\end{eqnarray}
\end{widetext}
where $\boldsymbol{\gamma} = \beta \boldsymbol{\alpha}$ is a Dirac
matrix.
The coefficients $A_l(Z,r)= A_l(Z) \, r/(r+0.07
Z^2\alpha^3)$ and 
$B_l(Z)$ are fitting factors, and in Ref. \cite{rad_pot} they were
found by fitting to the $5s$, $5p_{1/2}$, and $5p_{3/2}$ self-energy shifts for
hydrogen-like ions \cite{mohrkim}.
In that work, 
a single local potential was formed with
no dependence on the orbital angular momentum quantum number $l$.  
In the current work, we introduce an 
$l$-dependence to the potential in order to control the shifts to $d$ levels and
to improve the potential's accuracy for use in many-body calculations.

We keep the same fitting factors as those in Ref. \cite{rad_pot} for the $s$ and
$p$ levels, and for $d$ levels we introduce different factors, 
optimized to fit $5d_{3/2}$ and $5d_{5/2}$ self-energy shifts in
hydrogen-like ions \cite{mohrkim,highlse,shabaev},
\begin{widetext}
\begin{equation}
\begin{array}{ll}
\label{eq:AB}
A_l(Z) =
\Big\{
\begin{array}{l}
1.071-1.976x^2 -2.128x^3+0.169x^4\ ,\quad \\
0 \ , 
\end{array}
&
\begin{array}{l}
l = 0,\, 1\\
l  = 2 \ ,
\end{array}
\\
 B_l(Z) =
\Big\{
\begin{array}{l}
0.074+0.35Z\alpha\ , \\
0.056 + 0.050 Z \alpha + 0.195 Z^2 \alpha^2 \ ,
\end{array}
&
\begin{array}{l}
l = 0,\, 1 \\
l = 2 \ , 
\end{array}
\end{array}
\end{equation}
\end{widetext}
where $x=(Z-80)\alpha$.
For $l>2$, we set $A_l(Z)$ and $B_{l}(Z)$ to zero. 
The magnetic term is exact to first-order in $Z\alpha$, and no
fitting factors are introduced for it.

In Table \ref{tab:hlike} we present the self-energy shifts for 
hydrogen-like ions obtained using the point-nucleus radiative 
potential, Eqs. (\ref{eq:mag_point}), (\ref{eq:high_point}),
(\ref{eq:low_point}). The shifts $\delta \epsilon_{\rm SE}$ 
may be expressed as values of the function $F(Z\alpha)$ according 
to the relation \cite{mohrkim} 
\begin{equation}
\label{eq:fzalpha}
\delta \epsilon_{\rm SE}=\frac{\alpha}{\pi}\frac{(Z\alpha)^4}{n^3} F(Z\alpha)mc^2\ . 
\end{equation}
Our results for $F(Z\alpha)$ are tabulated alongside the  
results of exact self-energy calculations \cite{mohrkim,highlse,shabaev}.
The agreement for $n=5$ is particularly good, since the 
parameters of the radiative potential were found by fitting to these levels.
For $5s$, the deviations across all $Z$ are on the order of 0.1\%.
For other $s$ states presented in the table, the largest deviation is for Na $2s$,
where it is 2\%. For the $5p$ states, the deviation is typically on
the level of 1\%. However, around the nuclear charge for Rb, the 
self-energy shifts for $p_{1/2}$ transition from negative to
positive, and for Rb our radiative potential yields the wrong sign
for the shift, although the size of the shift is very small.
It is seen from Table \ref{tab:hlike} that the shifts for the $p$ waves with lower
principal quantum number $n$ deviate further from the exact
calculations, on the order of 10\%. 
For the $d$ levels considered, the radiative potential is typically accurate
to a few percent.
In Table \ref{tab:hlike} we have presented shifts for those states that we
consider to be of relevance in the study of neutral atoms. In
particular, valence level shifts for $s$, $p$, and $d$ waves, and the
shifts corresponding to the uppermost core $s$, $p$, and
$d$ waves, which affect the valence shifts through relaxation
effects. 

To obtain the finite-nucleus expressions for use in atomic codes, 
the point-nucleus expressions for the radiative potential are
folded with the nuclear density $\rho_{\rm nuc}$, 
\begin{equation}
V_{\rm SE}^{\rm fin}({\bf r})=\frac{1}{Z}\int d^3r' V_{\rm SE}^{\rm point}(|{\bf
    r}-{\bf r}'|) \rho_{\rm nuc} ({\bf r}') \ ,
\end{equation}
where the nuclear density is normalized as $\int \rho_{\rm nuc}({\bf r})d^3r=Z$. 
We find the following finite-nuclear-size expressions for 
the case of spherical symmetry of the nuclear density $\rho_{\rm nuc} ({\bf
  r})=\rho_{\rm nuc} (r)$,
\begin{widetext} 
\begin{eqnarray}
\label{eq:mag_rho}
V^{\rm fin}_{\rm mag}({\bf
  r})&=&\frac{i\alpha^3}{4}\boldsymbol{\gamma}\cdot\boldsymbol{\nabla}
\frac{1}{r}\int_0^{\infty}dr'
\int_1^{\infty}dt\frac{1}{t^3\sqrt{t^2-1}}\rho_{\rm nuc}(r')r'\Big[
(
e^{-2t|r-r'|/\alpha} -e^{-2t(r+r')/\alpha})-\frac{2t}{\alpha}(r+r'-|r-r'|)
\Big] \quad \\
&&\nonumber \\
\label{eq:high_rho}
V^{\rm fin}_{\rm high}(r) & = &
A_l(Z)\frac{\alpha}{r}\int_0^{\infty}dr' r' \rho_{\rm nuc} (r') \int_1^{\infty}dt\frac{1}{\sqrt{t^2-1}}\Big[ \Big(
1-\frac{1}{2t^2}\Big)[\ln(t^2-1)+4\ln(1/Z\alpha +1/2)] 
-\frac{3}{2}+\frac{1}{t^2}\Big] \times \nonumber \\
&&
\Big\{\frac{\alpha}{t}\Big(
e^{-2t(r+r')/\alpha} - e^{-2t|r-r'|/\alpha}\Big) 
+2r_A e^{2r_A t/\alpha}\Big(
E_1[(|r-r'|+r_A)2t/\alpha] -E_1[(r+r'+r_A)2t/\alpha] \Big)
\Big\} \quad \\
&&\nonumber \\
\label{eq:low_rho}
V^{\rm fin}_{\rm low}(r)&=&- B_l(Z) \frac{2\pi Z
  \alpha^3}{r}\int_0^{\infty}dr' r' \rho_{\rm nuc} (r') \Big[
(Z|r-r'|+1)e^{-Z|r-r'|}-(Z(r+r')+1)e^{-Z(r+r')}
\Big] 
\end{eqnarray}
\end{widetext}
where $E_1(x)=\int_x^{\infty}ds (e^{-s}/s)$ is the exponential integral and $r_A=0.07Z^2\alpha^3$.
We reduce these integrals further by considering the nucleus to be
modelled as a homogeneously charged sphere (step-function density),
\begin{widetext}
\begin{eqnarray}
\label{eq:mag_step}
&&V^{\rm step}_{\rm mag}({\bf r}) =
\Bigg\{
\begin{array}{ll}
\frac{3Zi}{\pi}\boldsymbol{\gamma}\cdot{\bf n}\int_1^\infty dt
\frac{1}{\varkappa^3\sqrt{t^2-1}}\Big(\frac{\alpha}{2tr}
\Big)^2\Big\{e^{-\varkappa}(1+\varkappa)
[\sinh(\frac{2tr}{\alpha})-\frac{2tr}{\alpha} \cosh(\frac{2tr}{\alpha})]
+\frac{1}{3}\big(\frac{2tr}{\alpha}\big)^3 
\Big\}\ , & \quad r\leq r_n \\
\frac{3Zi}{\pi}\boldsymbol{\gamma}\cdot{\bf n}\int_1^\infty dt
\frac{1}{\varkappa^3\sqrt{t^2-1}}\Big(\frac{\alpha}{2tr}
\Big)^2\Big\{e^{-2tr/\alpha}(1+2tr/\alpha)
[\sinh \varkappa -\varkappa\cosh \varkappa]
+\frac{1}{3}\varkappa^3 
\Big\}\ , & \quad  
r> r_n \ \ , 
\end{array} \\
&& \nonumber \\
\label{eq:high_step}
&&V^{\rm step}_{\rm high}(r) =
\Bigg\{
\begin{array}{ll}
-\frac{3}{2}A_l(Z) \frac{\alpha}{\pi}\frac{Z}{r}
\int_1^{\infty}dt I_1(t,Z)
\big\{ 
e^{-2tr/\alpha}\big( \frac{2}{\varkappa^3}\big)
[\varkappa \cosh \varkappa -\sinh \varkappa]
-\frac{r_A}{r_n^3} I_2(t,r,Z)
\big\}\ , & \  r\leq r_n \\
-\frac{3}{2} A_l(Z) \frac{\alpha}{\pi} \frac{Z}{r}\int_1^{\infty}dt
I_1(t,Z) \big\{
\frac{2}{\varkappa^3} \big[ \frac{r}{r_n}\varkappa -
e^{-\varkappa}(1+\varkappa)\sinh (2tr/\alpha) \big] 
-\frac{r_A}{r_n^3} I_2(t,r,Z)
\big\} \ , & \ 
r> r_n \ \ , 
\end{array} \\
&& \nonumber \\
\label{eq:low_step}
&&V^{\rm step}_{\rm low}(r) = - \frac{3}{2} B_l(Z) \frac{Z^2
  \alpha^3}{r_n^3 r}\int_0^{r_n}dr' r' \Big[
(Z|r-r'|+1)e^{-Z|r-r'|}-(Z(r+r')+1)e^{-Z(r+r')}
\Big] \ ,
\end{eqnarray}
\end{widetext}
where $r_n$ is the nuclear radius, $\varkappa=2tr_n/\alpha$, and 
\begin{widetext}
\begin{eqnarray}
I_1(t,Z)&=&\frac{1}{\sqrt{t^2-1}}\Big[
\Big(1-\frac{1}{2t^2}\Big)\Big[\ln(t^2-1)+4\ln\Big(\frac{1}{Z\alpha} +\frac{1}{2}\Big)\Big] 
-\frac{3}{2}+\frac{1}{t^2}\Big] \\
I_2(t,r,Z)&=&\int_0^{r_n}dr' r' e^{2r_A t/\alpha}\Big(
E_1[(|r-r'|+r_A)2t/\alpha] -E_1[(r+r'+r_A)2t/\alpha] \Big) \ .
\end{eqnarray}
\end{widetext}
This is the form of the radiative potential we use in subsequent
calculations, $V_{\rm SE}({\bf r})=V_{\rm SE}^{\rm step}({\bf r})$. 
The nuclear radii for the step-function density are found from the root-mean-square radii $r_{\rm rms}$ 
tabulated in Ref.~\cite{rms}, $r_n=\sqrt{5/3}\, r_{\rm rms}$. For
E119, we take $r_{\rm rms}=6.5\,{\rm fm}$, consistent with
Hartree-Fock-BCS theory \cite{BCS}. 
We have carried out numerical integration for the improper integrals
(integration over the variable $t$) using the GNU Scientific Library
routine QAGI \cite{gsl}.

We have checked the validity of the step-density approximation by performing
calculations for first-order self-energy shifts using a two-parameter
Fermi distribution for the nuclear density in 
Eqs. (\ref{eq:mag_rho}), (\ref{eq:high_rho}), (\ref{eq:low_rho}). We
took the 90\% to 10\% fall-off to be $2.3\, {\rm fm}$ for all atoms, 
and the half-density radius was found from this and from $r_{\rm rms}$
(above). We have found agreement to all digits presented, and therefore suggest the use 
of the simpler step-function form for the integrals. 
Energy shifts arising from the long-ranged $V_{\rm low}(r)$ are
insensitive to nuclear size, and the point-nucleus expression Eq. (\ref{eq:low_point})
may be used in place of Eq. (\ref{eq:low_step}). 
For the magnetic formfactor term $V_{\rm mag}({\bf r})$, differences
between the use of the point-nucleus and finite-nucleus expressions
appear in the energy shifts only for high $Z$.

\section{First-order shifts and comparison with exact QED}
 
In this section we calculate the first-order valence self-energy shifts in
different atomic potentials. Comparison of our results using
the radiative potential with results of exact self-energy calculations
performed in the same atomic potentials gives us a reliable indication 
of the accuracy of our approach.

The first-order shifts to the binding energies of the valence electron
are given by
\begin{equation}
\delta \epsilon_i^{(1)}=-\langle \varphi_i| V_{\rm SE}|\varphi_i\rangle \ .
\end{equation}
The orbitals are found from the solution of the relativistic equations
\begin{equation}
\label{eq:zeroth}
(c\boldsymbol{\alpha}\cdot {\bf p} +(\beta -1)c^2 -V_{\rm nuc}-V_{\rm el})\varphi_i= \epsilon_i\varphi_i\ ,
\end{equation}
where $\boldsymbol{\alpha}$ and $\beta$ are Dirac matrices, ${\bf p}$
is the momentum operator, and $V_{\rm nuc}$ and $V_{\rm el}$ are the
nuclear and electronic potentials. 
In our atomic structure calculations, we use a nuclear potential corresponding to a two-parameter Fermi
distribution for the nuclear density; the parameters are given in the final paragraph of the previous section.
We consider three
different electronic potentials in this paper: core-Hartree $V_{\rm CH}$,
Kohn-Sham $V_{\rm KS}$, and relativistic Hartree-Fock $V_{\rm HF}$. The first two are considered for
comparison with exact self-energy calculations performed in the same
atomic potentials (from Ref. \cite{sapcheng}) while the Hartree-Fock potential is the starting
point of our calculations in many-body perturbation theory. 
For more details about the core-Hartree and Kohn-Sham potentials, we
refer the reader to Ref. \cite{sapcheng}; for explicit expressions for
the relativistic Hartree-Fock potential, see Ref. \cite{johnson_book}.

We use the following form for the relativistic orbitals 
\begin{equation}
\varphi=\frac{1}{r}\Big(
\begin{array}{c}
f(r)\Omega_{\kappa m}\\
i\alpha g(r)\tilde\Omega_{\kappa  m}\\
\end{array}
\Big) \ ,
\end{equation}
where $f$ and $g$ are upper and lower radial components, 
the spherical spinor 
$\tilde\Omega_{\kappa m}=-(\boldsymbol{\sigma}\cdot {\bf
  n})\Omega_{\kappa m}=\Omega_{-\kappa m}$,
and the angular momentum quantum number 
$\kappa = \mp(j+1/2)$ for $j=l\pm 1/2$; $l$ is the orbital angular
momentum and $j$ the total angular momentum, with $m$ its projection on the 
$z$ axis.

The first-order valence shifts arising from the electric parts of the radiative potential are
\begin{eqnarray}
\delta\epsilon^{(1)}_{\rm high}&=&-\int_0^{\infty}drV_{\rm high}(f^2+\alpha^2 g^2) \\
\delta\epsilon^{(1)}_{\rm low}&=&-\int_0^{\infty}drV_{\rm low}(f^2+\alpha^2 g^2)
\end{eqnarray}
and the first-order shift from the magnetic formfactor is 
\begin{equation}
\label{eq:magfirst}
\delta\epsilon^{(1)}_{\rm mag}=-2\int_{0}^{\infty}dr fgH(r) \ ,
\end{equation}
where we have expressed the magnetic potential in terms of a
function $H(r)$ which we have defined such that 
\begin{equation}
\label{eq:Hdef}
V_{\rm mag}({\bf r})=i\boldsymbol{\gamma}\cdot {\bf n} H(r)/\alpha \ .
\end{equation}

\subsection{Core-Hartree and Kohn-Sham}
\label{sec:CHKS}

\begin{table*}
\caption{Breakdown of contributions 
to the first-order valence self-energy shift, using the FGRP method, to the Cs $6s$ binding energy in the core-Hartree
  approximation. Results for different nuclear approximations are given. Units: $10^{-5}\, {\rm a.u.}$ The final, screened
  finite-nucleus result $8.608\times 10^{-5}\, {\rm a.u.}$ corresponds to the value
  $F(Z\alpha)=0.01643$, presented in Table~\ref{tab:model}.}
\label{tab:breakdown_CH}
\begin{ruledtabular}
\begin{tabular}{lcccc}
Density approximation & $\delta \epsilon^{(1)}_{\rm mag}$ &$\delta \epsilon^{(1)}_{\rm high}$& $\delta \epsilon^{(1)}_{\rm low}$&$\delta \epsilon^{(1)}$ \\
\hline
Point-nucleus &1.391&6.583 &0.762 &8.736 \\
Step-function & 1.391 &6.577 &0.762 & 8.730 \\
Fermi distribution  & 1.391 &6.577 &0.762  & 8.730\\
Electron core &-0.013&-0.043&-0.065 &-0.121\\
Finite nucleus, screened &1.378&6.534&0.697&8.608\\
\end{tabular}
\end{ruledtabular}
\end{table*}

In Table \ref{tab:breakdown_CH} we present our core-Hartree results 
for Cs $6s$, with contributions from the magnetic and the electric components 
given separately. We present results corresponding to the use of different nuclear
approximations for the radiative potential: point-like, step-function density, and two-parameter Fermi distribution. For Cs and all other atoms considered 
in this work, we have found no difference, to all digits presented, between the use of the
step-function density (Eqs. (\ref{eq:mag_step}), (\ref{eq:high_step}), (\ref{eq:low_step})) and that of the Fermi
distribution (Eqs. (\ref{eq:mag_rho}), (\ref{eq:high_rho}),
(\ref{eq:low_rho})) with the same root-mean-square radius.

In calculations of exact QED in frozen atomic potentials, an effective charge is
used \cite{sapcheng}. This effective charge includes a screening of the nuclear charge
by the electrons. For comparison with exact QED, we should therefore
include this screening by the electrons. We do so in a simple manner by
replacing in Eqs. (\ref{eq:mag_rho}), (\ref{eq:high_rho}),
(\ref{eq:low_rho}) the nuclear density $\rho_{\rm nuc}$ with the
density of the electron core $\rho_{\rm el}(r)$ normalized such that $\int
\rho_{\rm el}(r) d^3r = -N_{\rm core}$, where $N_{\rm core}=Z-1$ is the number
of electrons in the core. This screening term for Cs $6s$ in the
core-Hartree approximation is given in the second last row of Table \ref{tab:breakdown_CH}.

\begin{table}[bh]
\caption{Finite-nucleus self-energy results $\delta \epsilon^{(1)}$ for neutral atoms in
  core-Hartree and Kohn-Sham potentials 
are presented, expressed as the function $F(Z\alpha)$. 
The results of this work (FGRP) are given alongside those of exact QED and
the model operator.}
\label{tab:model}
\begin{ruledtabular}
\begin{tabular}{lllllllll}
Atom & Z &State &\multicolumn{5}{c}{$F(Z\alpha)$}\\
&&& \multicolumn{3}{c}{core-Hartree} & \multicolumn{3}{c}{Kohn-Sham} \\
&&&FGRP&Exact\tablenotemark[1] &&FGRP& Exact\tablenotemark[1] & Model\tablenotemark[2]\\
\hline
Na &11&$3s$ &0.196&0.191&&0.185&0.181& 0.183\\
K&19 & $4s$ & 0.088 & 0.086 &&0.084 &0.083& 0.083 \\
Rb &37& $5s$ &0.0291  & 0.0286 && 0.0287& 0.0283& 0.0284\\
Cs &55& $6s$& 0.0164 & 0.0162&&0.0164&0.0162& 0.0163\\
Fr &87&$7s$& 0.0096 & 0.0096 &&0.0098& 0.0098& 0.0099\\
\end{tabular}
\end{ruledtabular}
\tablenotetext[1]{Sapirstein and Cheng, Ref. \cite{sapcheng}.}
\tablenotetext[2]{Shabaev {\it et al.}, Ref. \cite{shabaev}.}
\end{table}
 
In Table \ref{tab:model} we present our self-energy results for the
valence $s$ level shifts for Na through to Fr in the core-Hartree and
the Kohn-Sham approximations. Our results include electronic
screening, taken into account in the core-Hartree approximation in the 
manner described above. Our results are presented alongside the
exact self-energy shifts \cite{sapcheng} found in the same atomic
potentials. Results of the model operator approach of Shabaev {\it et al.}
\cite{shabaev} are shown also. It is remarkable how well the radiative 
potential approximates the exact self-energy in neutral atoms, at the 
level of about 1\%. E.g., for Na $3s$ in the Kohn-Sham approximation, 
we obtain the value $F=0.185$, while the exact value is $F=0.181$. The 
agreement is even better for the heavier atoms. 
The size of the deviation is only about twice the deviation seen
between the results of the model operator approach and the exact 
formalism.

\subsection{Hartree-Fock}

The relativistic Hartree-Fock approximation is the starting point 
for our treatment of many-body effects, as in this approximation
many-body perturbation theory in the residual Coulomb interaction 
is simplified; see Section \ref{sec:correlations}.
Therefore, we begin by considering the first-order valence self-energy corrections in the 
relativistic Hartree-Fock approximation. 

Note that in this and subsequent sections, we do not
include in our calculations the electronic density (or ``screening'') correction 
considered in Section~\ref{sec:CHKS}.

Our first-order results for the lowest 
$s$, $p$, and $d$ valence levels are presented in the fifth column 
of Table \ref{tab:all-order}. 
The $s$-level shifts range from about
$10^{-5}$~a.u. for Na to nearly $10^{-3}$~a.u. for E119. Of course, 
the $p$ and $d$ shifts are progressively smaller due to the
short-range nature of the self-energy interaction. While the 
shifts from the electric part of the potential lead to a reduction in
the binding energies for all states, the magnetic part of the
potential leads to shifts in the energies that are of opposite sign
for those levels with positive angular quantum number $\kappa$, 
i.e., for $p_{1/2}$ and $d_{3/2}$ waves. In some cases this leads 
to a delicate cancellation between the shifts arising from the magnetic formfactor
and from the low-frequency part of the electric formfactor. For
example, for Cs $6p_{1/2}$, the magnetic part of the potential
contributes $-0.2481 \times 10^{-5}~$a.u. and the low-frequency electric part of
the potential contributes $+0.2491 \times 10^{-5}~$a.u. 
The low-frequency part of the potential contains factors found by fitting to the hydrogen-like
$5p_{1/2}$ shift, and for this case there is also a degree of cancellation between the
terms (leaving a few percent of the size of one term). Since the
magnetic and electric terms are treated so differently in the atomic
structure calculations, 
a slight alteration in the orbitals can
produce a significant change in the shift of the level. This
limits the accuracy with which we can calculate the
Cs $6p_{1/2}$ shift. However, the size of this shift is small. 

In some cases, the magnetic shift dominates and the overall shift,
like the vacuum polarization contribution, is negative and leads to
increased binding.

\section{Core relaxation}

\begin{table*}
\caption{Contributions to the self-energy relaxation shifts $\delta
  \epsilon^{\rm relax}$ for Cs found using the FGRP method. First-order valence shifts,
  $\delta \epsilon^{(1)}$, are shown
  in column two for comparison. Contributions to relaxation shifts 
arising from self-energy corrections to individual
  core $s$ orbitals are given in columns 3 to 7,
  contributions from self-energy corrections to
  core $s$, core $p_{1/2}$, core $p_{3/2}$, core $d_{3/2}$, and
  core $d_{5/2}$ are presented in columns 8 to 12. 
The total relaxation shifts, $\delta \epsilon^{\rm relax}$, are given in
the final column. 
Units: ${\rm  a.u.}$}
\label{tab:deltaV}
\begin{ruledtabular}
\begin{tabular}{lcccccccccccc}
State &$\delta \epsilon^{(1)}$ &
\multicolumn{10}{c}{Contributions to the relaxation shift, $\delta
  \epsilon^{\rm relax}\times 10^{5}$}& $\delta \epsilon^{\rm relax}$\\ 
&$\times 10^{5}$&$1s$&$2s$&$3s$&$4s$&$5s$ & core $s$ & core $p_{1/2}$ & core
$p_{3/2}$& core $d_{3/2}$ & core $d_{5/2}$& $\times 10^5$ \\
\hline
$6s_{1/2}$&8.128&0.025&0.106&0.125&0.189&0.472&0.917&-0.064&-0.506&0.007&-0.014&0.341\\
$6p_{1/2}$&0.108&-0.030&-0.024&-0.021&-0.037&-0.218&-0.329&0.011&-0.170&0.005&-0.007&-0.491\\
$6p_{3/2}$&0.318&-0.018&-0.020&-0.015&-0.033&-0.242&-0.328&-0.024&-0.053&0.004&-0.009&-0.410\\
$5d_{3/2}$&-0.061&0.078&0.017&-0.036&-0.062&-0.732&-0.735&-0.086&-0.324&-0.011&0.005&-1.151\\
$5d_{5/2}$&0.072&0.073&0.019&-0.035&-0.064&-0.693&-0.700&-0.038&-0.461&-0.003&0.016&-1.186\\
\end{tabular}
\end{ruledtabular}
\end{table*}

In this and the following section we will consider how the account of 
many-body effects, in particular core relaxation and correlations
between the valence electron and the core,
influences the self-energy shifts in neutral
atoms. 

Core relaxation effects are found by adding the radiative potential to 
the potential felt by the core electrons, in this case the
Hartree-Fock potential of the core, $V_{\rm HF}+V_{\rm SE}$, and solving the 
relativistic equations (\ref{eq:zeroth}) for the core electrons
self-consistently. 
This leads to a new Hartree-Fock potential 
$V^{\rm SE}_{\rm HF}$. The correction to the zeroth-order Hartree-Fock
potential $V_{\rm HF}$ is 
$\delta V_{\rm HF}^{\rm SE}=V_{\rm HF}^{\rm SE}-V_{\rm HF}$; this may be
referred to as the relaxation correction to the potential.

The self-energy correction to the binding energy of the valence
electron may then be expressed as
\begin{equation}
\label{eq:relax_me}
\delta \epsilon_i=-\langle \varphi_i |V_{\rm SE}+\delta
V_{\rm HF}^{\rm SE}|\varphi_i\rangle =\delta
\epsilon_i^{(1)}+ \delta \epsilon_i^{\rm relax} \ .
\end{equation}
In actual calculations, however, we find the energies $\epsilon_i ^\prime $ from the solution of
the equation
\begin{equation}
\label{eq:relax1}
(c\boldsymbol{\alpha}\cdot {\bf p} +(\beta -1)c^2 -V_{\rm nuc}-V_{\rm
  SE} -V_{\rm HF}^{\rm SE})\varphi_i^\prime= \epsilon_i^\prime \varphi_i^\prime \ .
\end{equation}
The correction is given by
\begin{equation}
\label{eq:relax2}
\delta \epsilon_i= \epsilon^\prime_i - \epsilon_i \ .
\end{equation}
Note that energies from the solution of Eqs.~(\ref{eq:relax1}),
(\ref{eq:relax2}) 
include higher orders of the self-energy not contained in Eq.~(\ref{eq:relax_me}).

It is a simple matter to include the electric parts of the radiative potential into
the atomic structure codes. They may be added to the
Hartree-Fock or the nuclear
potential, as is done for Uehling
\cite{derev_relax,GBUehling}. Inclusion of the (off-diagonal) magnetic
part of the radiative potential is more involved, and the Dirac
equations must be modified accordingly,
\begin{eqnarray}
&\frac{df}{dr}= [ -\kappa/r+H(r)]f +[2+\alpha^2(\epsilon+V)]g&
\nonumber\\
&\frac{dg}{dr}=  -(\epsilon+V)f+[\kappa/r-H(r)]g \ , &
\end{eqnarray}  
where the magnetic term appears through the introduction of 
$H(r)$, defined by Eq. (\ref{eq:Hdef}). 
Here, the atomic potential includes the electric part of the radiative
potential, $V=V_{\rm nuc}+V_{\rm el}+V_{\rm high}+V_{\rm low}$.

In the sixth column of Table \ref{tab:all-order}, our self-energy
results with core relaxation included are presented. The corrections for $s$
levels enter at around 5\%. While the relaxation effect increases the 
size of the $s$-wave self-energy shift for the lighter atoms Na to Cs,
interestingly (unlike what we observed for the vacuum polarization \cite{GBUehling}) it
decreases the size of the shift for Fr and E119.

In Table \ref{tab:all-order} we present also the $s$-wave shifts 
calculated by Thierfelder and Schwerdtfeger \cite{thierfelder}. They performed
relativistic Hartree-Fock calculations with the radiative corrections
treated perturbatively; they used a modified version of the radiative
potential \cite{rad_pot}. 
There is good agreement between their results and our core-relaxed 
results.

For orbitals with $l>0$, account of the relaxation effect is absolutely
crucial for obtaining the correct sign and size of the shift, as was
seen for the vacuum polarization \cite{derev_relax,GBUehling}. While
the self-energy interaction is short-ranged, the relaxation potential 
$\delta V_{\rm HF}^{\rm SE}$ is long-ranged, making corrections to $p$ 
and $d$ waves significant. The trend in the sign and size of the 
shifts is less straightforward than what we observed for the 
very short-ranged Uehling potential \cite{GBUehling} due to the more 
complex form of the self-energy. It is not always 
clear from the start whether the shift will be positive or
negative.  The relaxation potential often produces 
a shift of the opposite sign as the first-order shift, and this may  
lead to a suppression of the first-order shift or a change in the sign
and magnitude of the shift.  
Moreover, as we mentioned in the previous section, for $p_{1/2}$ and
$d_{3/2}$ levels the different sign of the magnetic shift may produce
a high level of cancellation between terms. The change in the core 
potential may disturb this cancellation, making the 
shift relatively large.

The trend for the heavier elements is that the shifts for valence $d$
waves is larger than the shifts for $p$ waves, and that the size
of the $d$-wave shifts can be less than
an order of magnitude smaller 
than the $s$-wave shifts. For Cs, in the relaxed Hartree-Fock
approximation, the $5d$ shifts are only 7 to 8
times smaller than the $6s$ shift and of opposite sign.

Contributions to the relaxation shifts for Cs arising from the 
radiative corrections to individual $s$ states
of the core and from radiative corrections to core $p_{1/2}$, $p_{3/2}$, $d_{3/2}$ and
$d_{5/2}$ states are given in Table~\ref{tab:deltaV}. It is seen that
for all valence states, the relaxation shift comes mostly from the
radiative corrections to core $s$
states, with the uppermost core state giving the largest part of this
shift. The contributions from core $p$ states are also significant, 
with those from $d$ states small but non-negligible.

The relaxed self-energy shifts show a similar pattern overall with the 
relaxed vacuum polarization (Uehling) shifts \cite{GBUehling}.
The Uehling potential is extremely short-ranged, 
making the relative size of the relaxation correction for $d$ waves orders of magnitude 
larger than what we see for the self-energy. The relative
size of the relaxed $d$-wave shift to relaxed $s$-wave shift, however,
is roughly the same for the vacuum polarization and self-energy cases. We can understand this
by noting that most of the relaxation shift for $d$ waves (for the vacuum
polarization and the self-energy) arises from the radiative corrections to 
the uppermost $s$ wave of the core.

Indeed, much of the effect for
the $s$ and $d$ shifts can be determined by limiting the consideration
to the radiative shifts to $s$ waves only. The valence $s$-level shift
largely arises from the first-order valence shift; the valence $d$-level shift
arises largely from the radiative $s$-wave shift to the core. The
largest part of the self-energy shift to the $s$ waves comes from the 
high-frequency part of the electric formfactor term, and this term 
is rather similar in form to the Uehling potential. We understand 
this to be the reason for the comparable size of the ratio of the $d$- to
$s$-wave shifts for both vacuum polarization and self-energy.  

For the $p$ waves, 
the first-order valence shift and the shift arising from
the relaxation of the core are often close in magnitude and of opposite sign. 
Here, the radiative corrections to both $s$ and $p$ waves must be considered.

\section{Correlation corrections}
\label{sec:correlations}

\begin{table*}
\caption{First-order valence self-energy corrections to the binding energies
  for Cs in the FGRP approach, $\delta
  \epsilon^{(1)}_{\rm Br} = - \langle \varphi_{\rm Br} |V_{\rm
    SE}|\varphi_{\rm Br}\rangle$, where $\varphi_{\rm
    Br}$ is a solution of the Brueckner
  equation $(h_{\rm HF}+f_{\kappa}\Sigma_{\rm \kappa})\varphi_{\rm Br}
  =\epsilon\varphi_{\rm
  Br}$ and the correlation potential $\Sigma$ is the second-order
$\Sigma ^{(2)}$ or the all-orders $\Sigma ^{(\infty)}$. With no fitting,
$f_{\kappa}=1$ and $\epsilon=\epsilon_{\rm Br}$, while with fitting
$\epsilon=\epsilon_{\rm exp}$. 
The numbers in square brackets $[\,]$ denote powers of 10. Units: a.u.}
\label{tab:sigma}
\begin{ruledtabular}
\begin{tabular}{lccccccc}
State &$\epsilon_{\rm
  exp}$\tablenotemark[1]&\multicolumn{3}{c}{Second-order correlation
  potential $\Sigma^{(2)}$}&\multicolumn{3}{c}{All-orders correlation
  potential $\Sigma^{(\infty)}$}\\
 &&$\epsilon_{\rm Br}$&$\delta \epsilon^{(1)}_{\rm Br}$ &$\delta \epsilon^{(1)}_{\rm Br,fit}$ &
 $\epsilon_{\rm Br}$&$\delta \epsilon^{(1)}_{\rm Br}$ &$\delta \epsilon^{(1)}_{\rm Br,fit}$ \\
\hline
$6s_{1/2}$&-0.143098&-0.147671&1.225[-4]&1.134[-4] &-0.143262&1.120[-4]&1.118[-4]\\
$6p_{1/2}$ &-0.092166&-0.093578&1.708[-6]&1.588[-6]&-0.092436&1.610[-6] &1.588[-6]\\
$6p_{3/2}$&-0.089642&-0.090849&4.875[-6]&4.563[-6]&-0.089848&4.626[-6]&4.574[-6]\\
$5d_{3/2}$&-0.077035&-0.080029&-1.521[-6] \ &-1.353[-6] \ &-0.078015&-1.414[-6] \ &-1.359[-6] \ \\
$5d_{5/2}$&-0.076590&-0.079296&1.744[-6]&1.561[-6]&-0.077501& 1.629[-6]& 1.568[-6]\\
\end{tabular}
\end{ruledtabular}
\tablenotetext{Data from NIST, Ref. \cite{NIST}.}
\end{table*}

Use of the Hartree-Fock method as the starting approximation 
simplifies the perturbation theory in the residual Coulomb interaction, 
$1/|{\bf r}_i-{\bf r}_j|+V_{\rm HF}$, where $V_{\rm HF}$ is the
Hartree-Fock potential. The first non-vanishing correction for the 
interaction of the valence electron with the core appears in the 
second order in the Coulomb interaction. Diagrams of these 
second-order correlation corrections to the valence energies 
may be found in, e.g., Refs. \cite{review,dzuba}.

A non-local, energy-dependent potential 
$\Sigma^{(2)}({\bf r}_1,{\bf  r}_2,\epsilon_i)$  may be formed, defined 
such that its averaged value over the valence electron orbitals
corresponds to the correlation correction to the orbital energies,
$\delta \epsilon_i^{(2)}=\langle
\varphi_i|\Sigma^{(2)}({\bf r}_1,{\bf
  r}_2,\epsilon_i)|\varphi_i\rangle$. 
This potential is termed the ``correlation potential''. This
potential may be added to the relativistic Hartree-Fock 
equations for the valence electron, yielding so-called Brueckner 
orbitals $\varphi_{{\rm Br},i}$ and energies $\epsilon_{{\rm Br},i}$. 
This method takes into account the higher orders in $\Sigma$ in 
the Brueckner orbitals and energies.
The method of utilizing a potential to 
include correlations is called the {\it correlation potential method}.

Perturbation theory in the residual Coulomb interaction does not
converge rapidly; in some cases, the third-order corrections are
as large as the second \cite{DFSS}. Therefore, it is important to take
into account dominating classes of diagrams to all orders in the  
Coulomb interaction. A method for taking into 
account the higher orders of perturbation theory in the residual
Coulomb interaction was developed by Dzuba, Flambaum, and Sushkov
\cite{DFS1989}. In their method, the Coulomb lines in the second-order
correlation potential are modified to include an infinite
series of core polarization loops and an infinite series of
hole-particle interactions in those loops through the use of the 
Feynman diagram technique. In this case, the correlation correction 
to the energy may be expressed as $\delta \epsilon_i^{(\infty)}=\langle
\varphi_i|\Sigma^{(\infty)}({\bf r}_1,{\bf
  r}_2,\epsilon_i)|\varphi_i\rangle$. Again, this potential may 
be added to the relativistic Hartree-Fock equations for the valence 
electron to yield all-order Brueckner orbitals and energies.

The all-orders correlation potential method is a simple and effective 
approach that leads to some of the most accurate calculations of 
properties of heavy atoms, most notably for alkali atoms. 
One example is the atomic parity violating 
amplitude in Cs \cite{DFS1989APV,DFG2002,APV2012}. 

Inclusion of the correlation potential modifies the valence orbitals at large
distances, $r\gtrsim a_B$, pulling them towards the nucleus. 
This affects the form of the orbitals in the 
region where the self-energy interaction occurs through the 
normalization of the wave functions. See Ref. \cite{GBUehling} for 
more details about the correlation effects at small distances, and an 
illustration of the modification to the orbitals.

The second-order correlation potential $\Sigma^{(2)}$ is calculated 
using a B-spline basis set \cite{johnson} obtained by diagonalizing 
the relativistic Hartree-Fock operator on a
set of 40 splines of order $k=9$ within a cavity of radius $40\,{\rm a.u.}$ 
The exchange part of $\Sigma^{(\infty)}$ 
is also considered in the second-order, with (multipolarity-dependent)
factors used to screen the Coulomb interaction. 
For the direct part of $\Sigma^{(\infty)}$, the Feynman diagram technique is used for
inclusion of the core polarization and hole-particle classes of
diagrams; see Ref.~\cite{dzuba} for further details about the method. 

In the current work, we determine the effects on the self-energy 
shifts due to the use of the all-order Brueckner orbitals to take into account 
second and higher orders of perturbation theory in the residual
Coulomb interaction. We calculate these shifts for Cs, though 
we simplify the method for inclusion of third and higher orders of perturbation 
theory for the shifts to Na through to E119 by using a trivial fitting
procedure.
Inclusion of the effects of higher orders of perturbation theory
may be approximated simply by introducing factors before
$\Sigma^{(2)}$, with a different factor for each partial wave
$\kappa$. These factors are found by fitting the Brueckner energies to 
the experimental binding energies. The accuracy of calculations using
the all-orders $\Sigma^{(\infty)}$ may also be improved upon using
this method. 

In our recent work on the vacuum polarization shifts \cite{GBUehling}, 
we demonstrated the effectiveness of this fitting procedure for Cs. In 
Table \ref{tab:sigma} of the current work, we illustrate this approach 
for the case of the self-energy shifts. Indeed, while the bare 
second-order Brueckner results for the self-energy shifts differ 
from the all-order results in the second digit, the fitted
second-order results differ from the fitted all-order results in the 
third, or higher, digit. Therefore, we consider that the use of the fitted second-order
correlation potential for determining the valence-core correlations is
accurate to around 1\% or better.

In the seventh column of Table \ref{tab:all-order}, we present our
``final'' numbers for the self-energy shifts, taking into account both 
core relaxation and correlation corrections. The general trend in the 
effect of the Brueckner orbitals on the self-energy shifts is to
increase the size of the shift, and the largest corrections occur for 
the $d$-wave shifts which are typically enhanced by a factor of two or
more. This makes the relative size of the $d$ wave to $s$ wave shifts 
larger. For Cs, the self-energy shifts for the $5d$ levels are only
four to five times smaller than the $6s$ shift.

\begin{table*}
\caption{Self-energy corrections to binding energies in the FGRP
  approach. Experimental and 
zeroth-order relativistic Hartree-Fock binding 
energies are given in the third and fourth columns, respectively. First order valence 
  corrections $\delta \epsilon^{(1)} =-\langle \varphi |V_{\rm
    SE}|\varphi\rangle$ and shifts including 
core relaxation $\delta \epsilon$ are given in the following columns. 
The values in the seventh column correspond to the addition of fitted
$\Sigma^{(2)}$ (for E119, fitting factors from Fr are used) to the
relativistic Hartree-Fock equations; the
shift is found from the difference in energies when the self-energy is
included and excluded. In the final column the results of 
Thierfelder and Schwerdtfeger \cite{thierfelder} are presented for
comparison with our core-relaxed results $\delta \epsilon$.
The numbers in square brackets $[\,]$ denote powers of
10. Units: a.u.}
\label{tab:all-order}
\begin{ruledtabular}
\begin{tabular}{llcccccc}
Atom & State & $\epsilon_{\rm exp}$\tm[1] &$\epsilon_{\rm HF}$ & $\delta \epsilon^{(1)}$ & $\delta \epsilon$ & $\delta
\epsilon_{\rm Br, fit}$ & Other\tm[2] \\
\hline
Na & $3s_{1/2}$&-0.188858&-0.182033&1.068[-5]&1.125[-5]&1.275[-5]&1.118[-5]\\
&$3p_{1/2}$ &-0.111600& -0.109490& -5.698[-8] \ &-7.088[-7] \ &-8.162[-7] \ &\\
&$3p_{3/2}$ &-0.111521 &-0.109417&1.112[-7]&-4.882[-7] \ &-5.629[-7] \ &\\
&$3d_{3/2}$ &-0.055936&-0.055667&-2.644[-9] \ & 4.085[-11] &1.145[-10]&\\
&$3d_{5/2}$ &-0.055936&-0.055667 &1.770[-9]&3.407[-9]& 3.546[-9]&\\
K & $4s_{1/2}$ &-0.159516 &-0.147491&1.845[-5]& 1.974[-5]&2.543[-5]&2.013[-5]\\
&$4p_{1/2}$ &-0.100352&-0.095713&-1.023[-7] \ &-1.400[-6]&-1.805[-6] &\\
&$4p_{3/2}$ &-0.100089&-0.095498& 3.072[-7]&-8.411[-7]&-1.078[-6]&\\
&$3d_{3/2}$&-0.061387&-0.058067&-1.915[-8] \ &-2.568[-7]&-7.483[-7]&\\
&$3d_{5/2}$ &-0.061397& -0.058080&1.382[-8]&-2.647[-7] &-7.479[-7]&\\
Rb & $5s_{1/2}$ &-0.153507&-0.139291&4.836[-5]&\ 5.136[-5]  &\ 6.801[-5] &5.299[-5]\\
&$5p_{1/2}$ &-0.096193& -0.090816&1.381[-8]&-2.854[-6] & -3.806[-6] &\\
&$5p_{3/2}$ &-0.095110&-0.089986 &1.316[-6]&-1.083[-6] & -1.378[-6] &\\
&$4d_{3/2}$ &-0.065316&-0.059687&-1.098[-7] \ &-1.791[-6] & -4.355[-6] &\\
&$4d_{5/2}$ &-0.065318&-0.059745& 1.005[-7]& -1.790[-6] & -4.181[-6] &\\
Cs & $6s_{1/2}$& -0.143098&-0.127368&8.128[-5] &\ 8.431[-5]  &\ 1.152[-4]&8.735[-5]\\
&$6p_{1/2}$ &-0.092166& -0.085616 &1.077[-6] & -3.831[-6] & -5.355[-6] &\\
&$6p_{3/2}$ &-0.089642&-0.083785 &3.183[-6] &-9.203[-7]& -1.093[-6] &\\
&$5d_{3/2}$ & -0.077035&-0.064420& -6.066[-7]  \ & -1.212[-5]&-2.681[-5] &\\
&$5d_{5/2}$ &-0.076590&-0.064530&7.174[-7] & -1.115[-5] &-2.350[-5]  &\\
Fr & $7s_{1/2}$&-0.149670&-0.131076 &2.201[-4]&2.166[-4]&2.825[-4]&2.301[-4]\\
&$7p_{1/2}$ &-0.093913&-0.085911 &1.068[-5]& 1.276[-9]&1.959[-7] &\\
&$7p_{3/2}$ &-0.086228&-0.080443&1.034[-5]&-5.888[-8]&4.849[-7] &\\
&$6d_{3/2}$ &-0.075722&-0.062993&-8.046[-7]&-2.668[-5] &-5.991[-5] &\\
&$6d_{5/2}$ &-0.074812&-0.063444&1.968[-6]&-2.247[-5] &-4.528[-5] &\\
E119 & $8s_{1/2}$ &-&-0.152842&7.196[-4]&6.832[-4]&7.526[-4]&7.728[-4]\\ 
&$8p_{1/2}$ &-& -0.091697 &7.204[-5]&5.217[-5]&8.625[-5]&\\
&$8p_{3/2}$ &-&-0.075972&2.697[-5]&-2.001[-6]&-1.766[-6]&\\
&$7d_{3/2}$ &-&-0.061414& 4.523[-7] &-4.069[-5]&-1.068[-4]&\\
&$7d_{5/2}$ &-&-0.063000 & 4.717[-6] &-3.086[-5]&-6.009[-5]&\\
\end{tabular}
\end{ruledtabular}
\tablenotetext[1]{Data from NIST, Ref. \cite{NIST}.}
\tablenotetext[2]{Thierfelder and Schwerdtfeger \cite{thierfelder}, first-order perturbative treatment of the radiative
  potential in the relativistic Hartree-Fock approximation.}
\end{table*}

\section{Discussion}
\label{sec:discussion}

We expect that our self-energy shifts for $s$ and $d$ levels are
accurate to a few percent. The accuracy for at least some of the $p$
level shifts is lower due to the competing first-order and relaxation 
contributions and the competition between the magnetic 
and electric parts of the shift.

The accuracy of the radiative potential is limited by how well it 
reproduces the self-energy shifts for hydrogen-like ions. 
In future applications, a $\kappa$-dependence 
could be introduced to further improve the accuracy.
Indeed, a $\kappa$-dependent local potential is simple to implement
and introduces no additional complexity in the many-body methods. All
resulting wave functions remain orthogonal by virtue of the different
angular dependence. This may be contrasted with the $n$-dependence 
introduced into the potential in Ref. \cite{thierfelder} which 
brings about a non-orthogonality of the wave functions with different
principal quantum number $n$; the level of error introduced
through such non-orthogonality may be small, though should be checked
when used in many-body methods. 

Fitting the radiative potential to the self-energy shifts of individual hydrogen-like ions,
rather than fitting over the range $10\le Z\le 120$ simultaneously,
could help improve the accuracy for a specific atom or ion
under consideration. 
This would be the case, in particular, for those atoms or ions on the 
lighter or heavier sides of the range or those with lower principal
quantum number $n$, where the current deviations from the 
exact self-energy calculations for hydrogen-like ions are largest.

We should stress that 
the physical shift is the Lamb shift, well-approximated
by the one-loop self-energy and vacuum polarization shifts.
Typically, the self-energy shift is an order of magnitude larger than the vacuum
polarization shift and is of opposite sign. 
We have noticed a steeper increase of the vacuum polarization shifts \cite{GBUehling}
compared to the self-energy shifts with $Z$.  
There is a significant cancellation between the self-energy and 
Uehling contributions to the Lamb shift for the $8s$ and $8p_{1/2}$ levels. 
For the $8s$ shift, the self-energy contributes $7.526 \times
10^{-4}~{\rm a.u.}$ and the Uehling potential $-4.484\times
10^{-4}~{\rm a.u.}$, respectively. For the $8p_{1/2}$ shift, they
contribute $8.625\times 10^{-5}~{\rm a.u.}$
and $-7.643\times 10^{-5}~{\rm a.u.}$ (These values are taken from
Table IV of Ref. \cite{GBUehling} and from Table \ref{tab:all-order}
of the current work.) This may lead to a
suppression of the physical shift for these levels.

There are other contributions that will need to be considered for
a more accurate description of the Lamb shift, including account of electron 
screening (see Section \ref{sec:CHKS}), 
higher-orders in $Z \alpha$ vacuum polarization (Wichmann-Kroll), 
and higher-order loops.

Uncertainties in {\it ab initio} calculations of transition frequencies in alkali atoms are
at the level of 0.1\% \cite{dzuba},
roughly the level where the Lamb shifts enter. The accuracy is limited
by the incomplete account of electron-electron correlations. If the theoretical 
uncertainty can be reduced, then high-precision atomic studies of transition
frequencies, particularly involving $d$ levels, could provide 
a sensitive test of combined many-body and QED effects 
\cite{GBUehling}.

\section{Conclusion}

In this work we have studied the Flambaum-Ginges radiative
potential method in detail.
By calculating the self-energy
shifts in frozen atomic potentials and comparing with the results of
exact QED, we have shown that the accuracy of the method  
is high and comparable 
to that of the Shabaev-Tupitsyn-Yerokhin 
model operator approach \cite{shabaev}.

We have applied the radiative potential to the spectra of the series of alkali atoms
Na through to E119. 
We have demonstrated, through account of core relaxation and
valence-core electron correlations, that consideration of many-body effects is crucial 
for determining the correct sign and size of the shift for orbitals
with $l>0$ and for obtaining high accuracy for $s$ waves.  

Generally, the effect of the many-body corrections is to increase 
the size of the self-energy shifts. Remarkably, the many-body
enhancement is so large for the $d$-wave shifts that they 
approach the size of the shifts for $s$ waves.
High-precision atomic spectroscopic studies 
could provide a sensitive test of combined many-body and QED effects.

\section*{Acknowledgments}

This work was supported in part by the Australian Research Council, 
grant DE120100399.


\begin{thebibliography}{999}

\frenchspacing


\bibitem{wieman}

C. S. Wood, S. C. Bennett, D. Cho, B. P. Masterson, J. L. Roberts,
C. E. Tanner, and C. E. Wieman, Science {\bf 275}, 1759 (1997).

\bibitem{Sush_ueh}

O. P. Sushkov, Phys. Rev. A {\bf 63}, 042504 (2001).

\bibitem{JBS_ueh}

W. R. Johnson, I. Bednyakov, and G. Soff, Phys. Rev. Lett. {\bf 87}, 233001 (2001).

\bibitem{KF2002}

M. Yu. Kuchiev and V. V. Flambaum, Phys. Rev. Lett. {\bf 89}, 283002 (2002).

\bibitem{MST2002}

A. I. Milstein, O. P. Sushkov, and I. S. Terekhov,
Phys. Rev. Lett. {\bf 89}, 283003 (2002).

\bibitem{SPVC2003}

J. Sapirstein, K. Pachucki, A. Veitia, and K. T. Cheng, Phys. Rev. A
{\bf 67}, 052110 (2003).

\bibitem{SPTY2005}

V. M. Shabaev, K. Pachucki, I. I. Tupitsyn, and V. A. Yerokhin, 
Phys. Rev. Lett. {\bf 94}, 213002 (2005).

\bibitem{rad_pot}

V. V. Flambaum and J. S. M. Ginges, Phys. Rev. A {\bf 72}, 052115 (2005).

\bibitem{review}

J. S. M. Ginges and V. V. Flambaum, Phys. Rep. {\bf 397}, 63 (2004).

\bibitem{blundell}

S. A. Blundell, Phys. Rev. A {\bf 47}, 1790 (1993).

\bibitem{CCJS}

M. H.  Chen, K. T. Cheng,  W. R. Johnson, and J. Sapirstein, 
Phys. Rev. A {\bf 74}, 042510 (2006).

\bibitem{Schwerdt_review}

P. Schwerdtfeger, L. F. Pa\v{s}teka, A. Punnett, and P. O. Bowman,
Nucl. Phys. A, in press. 

\bibitem{mohr_review}

P. J. Mohr, G. Plunien, and G. Soff, Phys. Rep. {\bf 293}, 227 (1998).

\bibitem{labzowsky}

 L. Labzowsky, I. Goidenko, M. Tokman, and P. Pyykk\"o, Phys. Rev. A
 59, 2707 (1999).

\bibitem{sapcheng}

J. Sapirstein and K. T. Cheng, Phys. Rev. A {\bf 66}, 042501 (2002).

\bibitem{pyykko}

P. Pyykk\"o and L.-B. Zhao, J. Phys. B {\bf 36}, 1469 (2003).

\bibitem{roberts}

B. M. Roberts, V. A. Dzuba, and V. V. Flambaum, Phys. Rev. A {\bf 87},
054502 (2013).

\bibitem{DDFG}

T. H. Dinh, V. A. Dzuba, V. V. Flambaum, and J. S. M. Ginges, 
Phys. Rev. A {\bf 78}, 022507 (2008).

\bibitem{radium}

J. S. M. Ginges and V. A. Dzuba, Phys. Rev. A {\bf 91}, 042505 (2015).

\bibitem{nature}

T. K. Sato {\it et al.}, Nature {\bf 520}, 209 (2015).

\bibitem{nandysahoo}

D. K. Nandy and B. K. Sahoo, Phys. Rev. A {\bf 88}, 052512 (2013).

\bibitem{thierfelder}

C. Thierfelder and P. Schwerdtfeger, Phys. Rev. A {\bf 82}, 062503 (2010).

\bibitem{chantler}

 J. A. Lowe, C. T. Chantler, and I. P. Grant, Rad. Phys. Chem. {\bf 85}, 118 (2013).

\bibitem{IGD}

P. Indelicato, O. Gorceix, and J. P. Desclaux, J. Phys. B {\bf 20},
651 (1987).

\bibitem{welton}

T. A. Welton, Phys. Rev. {\bf 74}, 1157 (1948).

\bibitem{tupbers}

I. I. Tupitsyn and E. V. Berseneva, Opt. Spectrosc. {\bf 114}, 682 (2013).

\bibitem{shabaev}

V. M. Shabaev, I. I. Tupitsyn, and V. A. Yerokhin, Phys. Rev. A {\bf 88}, 012513 (2013).

\bibitem{dyall}

K. G. Dyall, J. Chem. Phys. {\bf 139}, 021103 (2013).

\bibitem{QEDMOD}

V. M. Shabaev, I. I. Tupitsyn, V. A. Yerokhin, Comp. Phys. Comm. {\bf 189}, 175 (2015).

\bibitem{GBUehling}

J. S. M. Ginges and J. C. Berengut, accepted to J. Phys. B; online preprint arxiv:1511.01459 (2015).

\bibitem{pyykkoRatio}

P. Pyykk\"o, M. Tokman, and L. N. Labzowsky, Phys. Rev. A {\bf 57},
R689 (1998).

\bibitem{ID1990}

P. Indelicato and J. P. Desclaux, Phys. Rev. A {\bf 42}, 5139 (1990).

\bibitem{derev_relax}

A. Derevianko, B. Ravaine, and W. R. Johnson, Phys. Rev. A {\bf 69},
054502 (2004).

\bibitem{mohrkim}

P. J. Mohr and Y.-K. Kim, Phys. Rev. A {\bf 45}, 2727 (1992).

\bibitem{highlse}

{\'E}.-O. Le Bigot, P. Indelicato, and P. J. Mohr, Phys. Rev. A {\bf 64}, 052508 (2001).

\bibitem{rms}

I. Angeli and K. P. Marinova, At. Data Nucl. Data Tables {\bf 99}, 69 (2013).

\bibitem{BCS}

S. Goriely, F. Tondeur, and J. Pearson, At. Data Nucl. Data Tables
{\bf 77}, 311 (2001).

\bibitem{gsl}

M. Galassi {\it et al.}, GNU Scientific Library Reference Manual (3rd
Ed.); http://www.gnu.org/software/gsl/.

\bibitem{johnson_book}

W. R. Johnson, Atomic Structure Theory: Lectures on Atomic Physics
(Springer, Berlin, 2007). 

\bibitem{dzuba}

V. A. Dzuba, Phys. Rev. A {\bf 78}, 042502 (2008).

\bibitem{DFSS}

V. A. Dzuba, V. V. Flambaum, P. G. Silvestrov, and O. P. Sushkov,
Phys. Lett. A {\bf 131}, 461 (1988).

\bibitem{DFS1989}

V. A. Dzuba, V. V. Flambaum, and O. P. Sushkov, Phys. Lett. A {\bf 140}, 493 (1989). 

\bibitem{DFS1989APV}

V. A. Dzuba, V. V. Flambaum, O. P. Sushkov, Phys. Lett. A {\bf 141}, 147 (1989).

\bibitem{DFG2002}

V. A. Dzuba, V. V. Flambaum, and J. S. M. Ginges, Phys. Rev. D {\bf 66}, 076013 (2002).

\bibitem{APV2012}

V. A. Dzuba, J. C. Berengut, V. V. Flambaum, and B. Roberts,
Phys. Rev. Lett. {\bf 109}, 203003 (2012).

\bibitem{johnson}

W. R. Johnson and J. Sapirstein, Phys. Rev. Lett. {\bf 57}, 1126 (1986).

\bibitem{NIST}

A. Kramida, Yu. Ralchenko, J. Reader, and NIST ASD Team (2014). NIST
Atomic Spectra Database (ver. 5.2), [Online]. Available:
http://physics.nist.gov/asd. National Institute of Standards and Technology, Gaithersburg, MD. 

\end{thebibliography}
\end{document}